\documentclass[12pt]{iopart}

\usepackage{graphicx}
\usepackage{iopams}
\DeclareTextSymbol{\degre}{T1}{6}
\DeclareTextSymbol{\degre}{OT1}{23}
\newcommand{\Xh}{\ensuremath{\chi_\subs{hom}}}
\newcommand{\Xl}{\ensuremath{\chi_\subs{line}}}
\newcommand{\Xt}{\ensuremath{\chi_\subs{tot}}}

\newcommand{\subs}[1]{{\mbox{\tiny #1}}}
\newcommand{\Velec}{\ensuremath{v_\subs{el}}}

\begin{document}

\title[Field test of a continuous-variable QKD prototype]{Field test of a continuous-variable quantum key distribution prototype}

\author{S~Fossier$^{1,2}$, E~Diamanti$^2$, T~Debuisschert$^1$, A~Villing$^2$, R~Tualle-Brouri$^2$ and P~Grangier$^2$}
\address{$^1$Thales Research \& Technology France, RD 128, 91767 Palaiseau Cedex, France}
\address{$^2$Laboratoire Charles Fabry de l'Institut d'Optique -- CNRS -- Univ. Paris-Sud, Campus Polytechnique, RD 128, 91127 Palaiseau Cedex, France}

\ead{simon.fossier@institutoptique.fr}

\begin{abstract}
We have designed and realized a prototype that implements a continuous-variable quantum key distribution protocol based on coherent states and reverse reconciliation. The system uses time and polarization multiplexing for optimal transmission and detection of the signal and phase reference, and employs sophisticated error-correction codes for reconciliation. The security of the system is guaranteed against general coherent eavesdropping attacks. The performance of the prototype was tested over preinstalled optical fibres as part of a quantum cryptography network combining different quantum key distribution technologies. The stable and automatic operation of the prototype over 57 hours yielded an average secret key distribution rate of 8~kbit/s over a 3~dB loss optical fibre, including the key extraction process and all quantum and classical communication. This system is therefore ideal for securing communications in metropolitan size networks with high speed requirements.

\end{abstract}


\maketitle

\section{Introduction}
\label{sec:intro}

Experimental quantum key distribution (QKD) has been the subject of intense research efforts during the last decade; these efforts have led to an impressive progress, which allowed QKD to become the field in quantum information processing that is the closest to applications. The majority of the QKD systems that have been realized implement discrete-variable or distributed-phase-reference protocols~\cite{scarani:quantph2008}, which use properties of single photons to encode key information. Continuous-variable (CV) QKD protocols, in which light carries continuous information such as the value of the quadrature of a coherent state, have been proposed as another option, which opens the door to very high secret key generation rates~\cite{ralph:pra1999,hillery:pra2000,cerf:pra2001,silberhorn:prl2002,grosshans:prl2002,grosshans:nature2003,weedbrook:prl2004,heid:pra2007}. In the protocol that we have implemented~\cite{grosshans:nature2003}, the key information is encoded by Alice in two orthogonal quadratures $X$ and $P$ (or equivalently the amplitude and phase) of a train of coherent states that are modulated according to a centred bi-Gaussian distribution. These states are sent along with a phase reference through the quantum channel to Bob, who randomly measures one of the two quadratures using a homodyne detection setup. Alice and Bob then extract a secret common binary key from their data, by performing classical procedures for channel characterization, error correction and privacy amplification. The security of this protocol stems from Heisenberg inequalities satisfied by the quantum continuous variables accessible to Alice, Bob, and the eavesdropper Eve. Security proofs of this protocol against general individual and collective eavesdropping attacks have been provided~\cite{grosshans:nature2003,grosshans:prl2004,navasques:prl2006,garcia-patron:prl2006}, and have been recently completed by unconditional security proofs against the most general type of attacks~\cite{renner:quantph2008}.

From a practical point of view, the coherent-state CVQKD protocol requires a simple system architecture and eliminates the need for specific resources such as single-photon sources and detectors. It was first implemented in a proof-of-principle table-top experiment at near-infrared wavelength (780 nm), based on a pulsed, shot-noise limited homodyne detector~\cite{grosshans:nature2003}. Subsequently, the system went through successive phases of development that included the implementation of advanced error-correction algorithms and the operation at infrared wavelength (1550 nm) in order to enhance the performance of the system using standard, fast and efficient products of the telecommunication industry~\cite{lodewyck:pra2005,lodewyck:pra2007}. The experimental implementation of a partial intercept-and-resend eavesdropping attack on such a system confirmed the entanglement-breaking bound for the coherent-state CVQKD protocol through a direct measurement of the system's excess noise~\cite{lodewyck:prl2007}. However, despite this progress, the developed systems remained laboratory experimental setups, unsuitable for implementation of quantum key distribution in standard fibre optic networks. Here, we present a portable CVQKD prototype that provides stable and automatic distribution of secret keys over several days at high rates. To reach this goal, several experimental advancements were required, such as the time and polarization multiplexing of the signal and phase reference in the quantum channel, as well as the implementation of automation and hardware control procedures, combined with advanced reconciliation and key verification techniques. The prototype was tested in a field implementation of a quantum cryptography network developed within the European Integrated Project SECOQC~\cite{secoqc}, which brought together prototypes based on various quantum key distribution technologies~\cite{yuan:apl2008,poppe:optexp2004,stucki:apl2005,idquantique}. The average secret key distribution rate provided by the CVQKD prototype over 3 days and through a 3~dB channel (corresponding to a 15~km standard optical fibre) was 8~kbit/s, including all quantum and classical communication. This system is therefore most suitable for use in metropolitan-size secure networks that require high communication rates.

\section{CVQKD prototype layout}
\label{sec:layout}

The optical layout of the CVQKD prototype we have designed is shown in figure~\ref{fig:setup}. The setups of Alice and Bob are entirely composed of fibre optic components operating at telecom wavelength, pigtailed with polarization-maintaining (PM) single-mode fibres. They have been designed to implement the protocol briefly discussed in the introduction and presented in detail in~\cite{grosshans:nature2003}.

\subsection{Optical setup}
\label{sec:setup}

\begin{figure}[ht]
  \centering
  \includegraphics[width=\columnwidth]{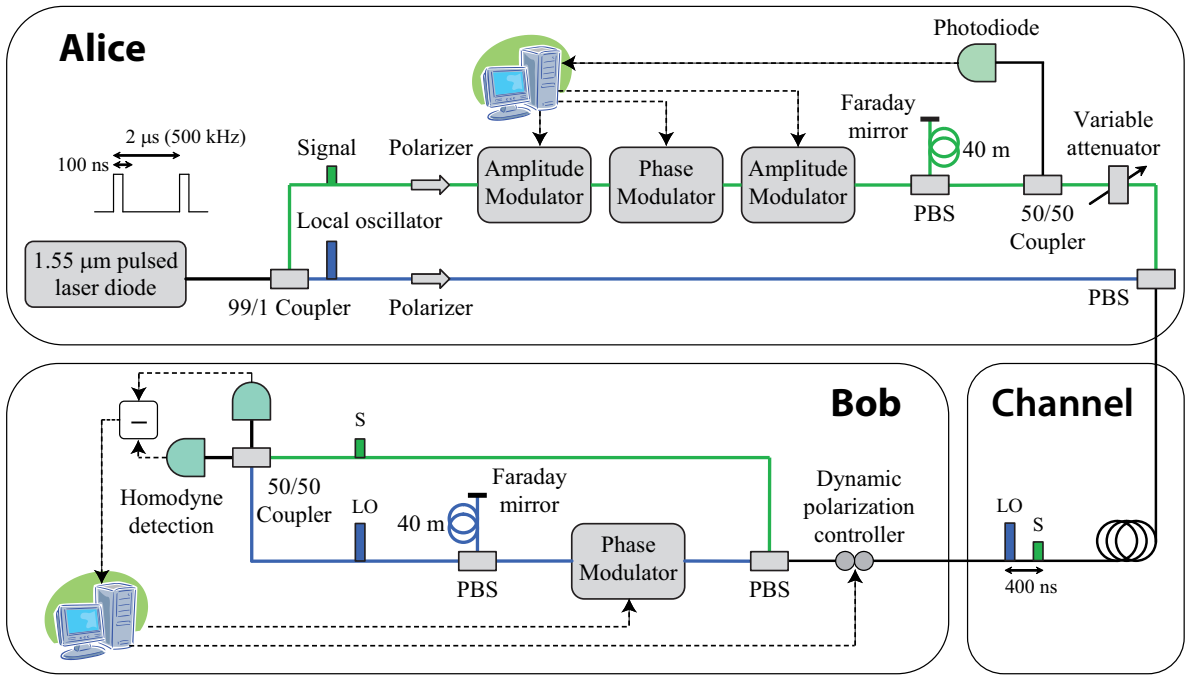}
\caption{Optical layout of the CVQKD prototype.}
\label{fig:setup}
\end{figure}

As shown in figure~\ref{fig:setup}, Alice generates coherent light pulses using a 1550 nm telecom laser diode pulsed with a frequency of 500 kHz. The length of the generated pulses is 100 ns. The pulses are then separated into a weak signal and a strong local oscillator (LO) using a highly asymmetric coupler. The signal is randomly modulated following a centred Gaussian distribution in both quadratures, using amplitude and phase modulators. In order to ensure the randomness of the modulation, a Quantis true random number generator has been implemented in the system. The mean intensity of the pulses is then adjusted roughly by a variable attenuator and finely by a second amplitude modulator, so that the variance of the Gaussian distribution reaches a target value of $V_\subs{A} N_0$, where $N_0$ is the shot noise variance.

Time and polarization multiplexing are used so that the signal and LO are transmitted to Bob in the same optical fibre without interfering. For the time multiplexing, two identical 400 ns delay lines are inserted into the system, one in Alice's signal path and one in Bob's LO path, as shown in figure~\ref{fig:setup}. Each line is composed of a 40~m non-PM single-mode fibre followed by a Faraday mirror. The Faraday mirror is a non-reciprocal optical device composed of a standard mirror and a 45$^\circ$ Faraday rotator; it therefore reflects the pulse by imposing a 90$^\circ$ rotation on its polarization. This system practically eliminates all birefringence-induced polarization drifts that the pulses experience during propagation in the delay line. Furthermore, to achieve polarization multiplexing, the pulses are coupled in the transmission fibre using a polarization beam splitter (PBS)~\cite{marand:ol1995,qi:pra2007}. Therefore, the signal and LO pulses propagate through the quantum channel with orthogonal polarizations, and they are also delayed in time. With this configuration and provided that the quantum channel features a sufficiently low crosstalk between orthogonal polarizations, which is indeed less than -30 dB in our case, the two pulses can be demultiplexed at Bob's site very efficiently and with minimal losses, as shown in figure~\ref{fig:setup}.

Finally, in Bob's system, the signal and LO interfere in a pulsed, shot-noise limited homodyne detector. This detection system outputs an electric signal, whose intensity is proportional to the quadrature $X_\phi$ of the signal, where $\phi$ is the phase difference between the signal and the LO. Following the implemented protocol, Bob measures randomly either $X_0$ or $X_{\pi/2}$ to select one of the two quadratures. For this purpose, he imposes randomly a $\pi/2$ phase shift to the local oscillator using a phase modulator placed in the LO path.

\subsection{Feedback control and automation procedures}
\label{sec:automation}

To ensure the automated operation of the described setup, we implemented several feedback control procedures to eliminate the effect of polarization and temperature drifts. More specifically, a dynamic polarization controller, placed at the output of the quantum channel as shown in figure~\ref{fig:setup}, is used to correct the polarization drifts that occur in the transmission fibre because of temperature changes or mechanical strains. During system initialization, the controller explores randomly the Poincar\'e sphere to find an optimal polarization state at the output of the channel, and subsequently adjusts in real time this state to compensate for all such drifts.

Another implemented procedure is linked to the fact that the active material used in the amplitude and phase modulators, namely lithium niobate (LiNbO$_3$), is very sensitive to temperature changes. As a consequence, the voltages that need to be applied to reach target modulation values constantly drift with temperature. To overcome this problem, an automation software measures and corrects in real time these drifts, by analyzing the outputs of the three photodiodes present in the system. In particular, the photodiode placed in Alice's signal path is used for the feedback control of the amplitude modulators, while the interference signal in the homodyne detector yields information on the relative phase between the signal and LO pulses, which is then used to control the phase modulators.

The time required for a complete feedback control of the system is 1~minute, which is small compared to typical temperature drifts and allows the devices to remain in an optimal state over the entire system operation time.

\section{Security of the CVQKD prototype}
\label{sec:paranoid}

The security against individual Gaussian eavesdropping attacks of the coherent-state CVQKD protocol that we have implemented with reverse reconciliation was first proven in~\cite{grosshans:nature2003}, and the proof was later extended to general individual attacks~\cite{grosshans:prl2004}. This was followed by security proofs against general collective attacks~\cite{navasques:prl2006,garcia-patron:prl2006}. Recently, the unconditional security of this protocol against coherent attacks, the most general attacks allowed by quantum mechanics, has also been proven~\cite{renner:quantph2008}. Several of these proofs make the assumption that all the losses and detection-added noise present in Bob's apparatus are available to the eavesdropper. In a more realistic setting, however, Eve does not have access to Bob's system~\cite{grosshans:nature2003,lodewyck:pra2007}. For the practical prototype we have implemented, we have therefore taken into account this ``realistic'' assumption.

\begin{figure}[b]
  \centering
  \includegraphics[width=\columnwidth]{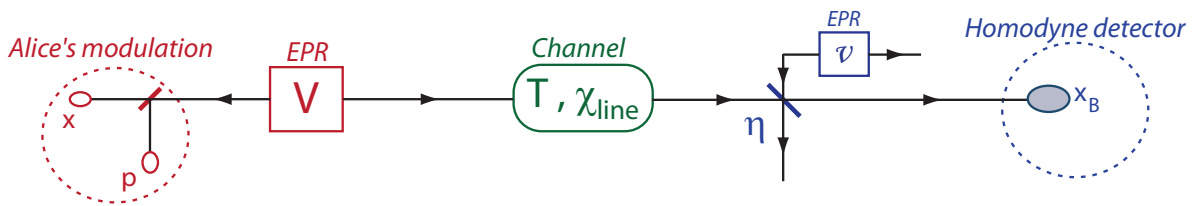}
\caption{Schematic representation of the entanglement-based description of the CVQKD protocol.}
\label{fig:EB}
\end{figure}

In the following, we review the results for security against general collective attacks when Eve does not have access to Bob's setup and when Alice and Bob use Bob's data to extract the final secret key (reverse reconciliation)~\cite{grosshans:nature2003}. Since coherent attacks were shown not to be more powerful than collective attacks for the implemented protocol~\cite{renner:quantph2008}, the following expressions remain valid for this case as well, guaranteeing the unconditional security of the corresponding CVQKD system.

The security proofs against collective attacks rely on an entanglement-based description of the protocol, shown in figure~\ref{fig:EB}, which is formally equivalent to the prepare-and-measure scheme presented in the previous section. In this scheme, Alice's bi-Gaussian modulation is modeled by an EPR state of variance $V$, on one half of which Alice performs a measurement of both quadratures. Bob's detector noise and losses are modeled by a beamsplitter, one input of which is the output state of the quantum channel and the other one half of an EPR state of appropriate variance. Based on this description, the final secret information available to Alice and Bob is given by the expression~\cite{lodewyck:pra2007}:
\begin{eqnarray}
\Delta I &= \beta\,I_\subs{AB} - \chi_\subs{BE}
\label{eq:rate}
\end{eqnarray}
where $\beta$ is the reconciliation efficiency, discussed in section~\ref{sec:time}, and: \begin{eqnarray*}
 I_\subs{AB} &= \frac{1}{2}\log_2\frac{V+\Xt}
 {1+\Xt}\\
\chi_\subs{BE} &= G\left(\frac{\lambda_1-1}{2}\right) +
	G\left(\frac{\lambda_2-1}{2}\right)	 - G\left(\frac{\lambda_3-1}{2}\right) -
	G\left(\frac{\lambda_4-1}{2}\right)
\end{eqnarray*}
with \begin{eqnarray*}
	G(x)=(x+1)\log_2(x+1)-x\log_2x\qquad &\Xt=\Xl+\Xh/T\\
	\Xl=1/T-1+\varepsilon\qquad&\Xh=(1+\Velec)/\eta-1\\
	\lambda^{2}_{1,2}=\frac{1}{2}(A \pm \sqrt{A^2-4B}) \qquad &\lambda^2_{3,4} = \frac{1}{2}(C\pm\sqrt{C^2-4D})\\
	A  =  V^2(1-2T)+2T+T^2(V+\Xl)^2\qquad
	&B  =  T^2 (V\Xl+1)^2\\
	C  =  \frac{V\sqrt{B}+ T(V+\Xl) + A \Xh}{T(V+\Xt)} \qquad
	&D  =  \sqrt{B}\frac{V + \sqrt{B} \Xh }{T(V+\Xt)}
\end{eqnarray*}
\\
\indent In the above expressions, $V_\subs{A} = V-1$ is Alice's modulation variance at the input of the channel, expressed in shot noise units, $T$ is the transmission efficiency of the quantum channel, $\varepsilon$ is the excess noise at the channel's input, $\eta$ is the global transmission efficiency of Bob's apparatus, $\Velec$ is the noise (mostly electronic) at Bob's setup input, and $\lambda_{1,2,3,4} \geq 1$. Given the parameters $V_\subs{A}$, $T$, $\varepsilon$, $\eta$ and $\Velec$, Alice and Bob can therefore calculate the information they share after the quantum communication, $I_\subs{AB}$, as well the maximal bound on the information available to the eavesdropper, $\chi_\subs{BE}$. They can then derive from equation~\ref{eq:rate} the maximal amount of secret information they can extract from their continuous data to form the secret key.

The security provided by the CVQKD prototype that we have implemented corresponds to the above analysis and, in the following sections, the theoretical values for the secret key generation rate are derived using equation~\ref{eq:rate}.

\section{From continuous data to a secret key}
\label{sec:reconciliation}

\subsection{General scheme}
\label{sec:scheme}

In the previous sections, we have discussed in detail the quantum transmission phase of the coherent-state CVQKD protocol. In continuous-variable QKD, however, a significant amount of data post-processing is also required, which presents important differences compared to discrete-variable or distributed-phase-reference QKD schemes. In those protocols, when an ideal system with no detector dark noise is considered, and when no eavesdropper is present, the error rate is zero, hence the sifted data of Alice and Bob are identical. On the other hand, in CVQKD, even with a noiseless detector and no eavesdropping, Bob's measurements are always affected by Heisenberg uncertainty relations, which result in a fundamental quantum noise, the so-called shot noise, added to every quadrature measurement. Therefore, after quantum transmission, Alice and Bob do not share identical quadrature values, but only correlated data.

\begin{figure}[t]
  \centering
  \includegraphics[width=.7\columnwidth]{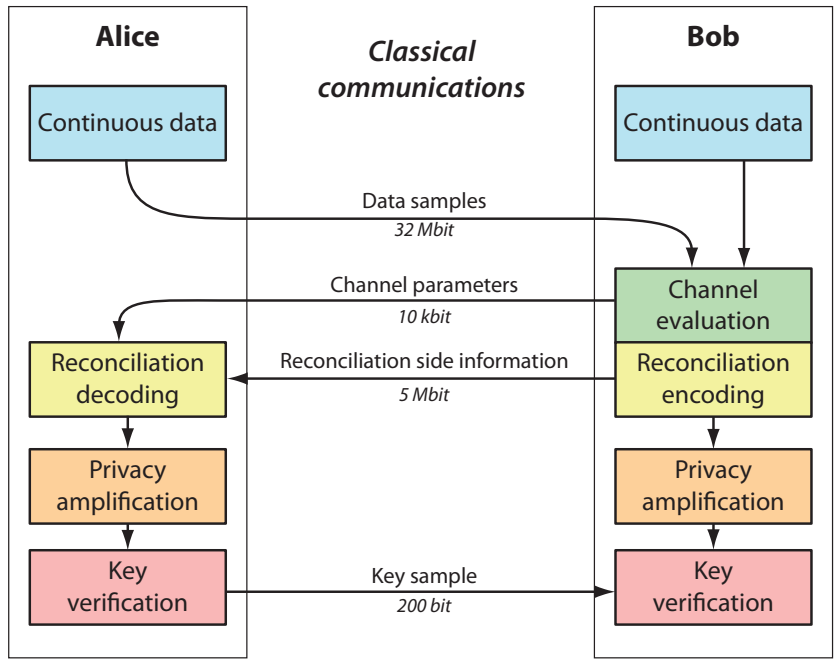}
\caption{Block description of the data post-processing steps. The size of the data sent between Alice and Bob on the classical channel is expressed in number of transmitted bits per final key. In the current implementation, 2 million pulses are used for each key generation. The amplitude and phase (coded on 16 bits) of 1 million of them are sent for the channel evaluation, and the remaining 1 million are the material used for the key generation. With the parameters given in section~\ref{sec:labo}, the final key size is typically of the order of 150\,000 bits, for a total processing time of 20 seconds.}
\label{fig:Rec}
\end{figure}

The various steps performed by Alice and Bob to convert their partially secret correlated continuous data into perfectly secure identical discrete data, i.e. a secret key, are summarized in figure~\ref{fig:Rec}. First, Alice and Bob need to evaluate the parameters of the transmission. Alice sends for this purpose random samples of her data to Bob, who compares them with his data. In this way, the parameters $V_\subs{A}$, $T$ and $\varepsilon$ that appear in the calculations are evaluated. The parameters $\Velec$ and $\eta$ are determined before the transmission by a calibration of Bob's apparatus. Using equation~\ref{eq:rate}, Alice and Bob can then estimate the secret information $\Delta I$ present in the shared data.

Subsequently, Alice and Bob apply an error-correction (or reconciliation) algorithm to their data. In particular, in the prototype that we have implemented, they first discretize their respective continuous data by dividing the Gaussian distribution of each quadrature in 16 slots, and attributing to each continuous value a 4-bit label corresponding to its position in the distribution. After this process, in order for Alice to correct the errors that appear in her data with respect to Bob's data (reverse reconciliation), the two parties perform a multilevel reconciliation process based on Low Density Parity Check (LDPC) codes, which has been described in detail in~\cite{lodewyck:pra2007}. This process has a finite efficiency, which plays a crucial role in the performance of the CVQKD system, as we will see in section~\ref{sec:time}. The reconciliation process also inevitably reveals to the eavesdropper a certain amount of side information, which has to be taken into account during the next phases of the data processing.

After reconciliation, Alice and Bob share identical discrete data, which are only partially secret. To extract the secret information present in this data, Alice and Bob use privacy amplification algorithms based on hash functions that combine the information known to Eve with the information she ignores, yielding at the end of the process a shorter bit sequence entirely unknown to her. This output sequence is the perfectly secure key shared by Alice and Bob. Note that, for computational speed reasons, we use a combination of two hash functions chosen respectively from a non-universal family that allows rapid computation, and from a universal family that is slower to process. As described in~\cite{lodewyck:pra2007}, using a correct combination of these two families allows in practice the same security parameters as using a single universal family.

The last step of the data post-processing procedure shown in figure~\ref{fig:Rec}, namely key verification, is described below.

\subsection{Integration into a real-size network and real-time operation requirements}
\label{sec:integration}

The CVQKD prototype has been specifically developed to satisfy the requirements of the European Integrated project SECOQC~\cite{secoqc}. This project aimed at developing a metropolitan-size QKD network implemented on preinstalled optical fibres, and led to a real-size demonstration of such a network in October 2008, on the Siemens fibre network of Vienna. The fundamental communication layer of this network, called ``quantum backbone'', is composed of seven QKD links, based on five different technologies \cite{lodewyck:pra2007,yuan:apl2008,poppe:optexp2004,stucki:apl2005,idquantique}. The systems of this layer distribute keys between two points of the network; these keys can therefore only be used for a point-to-point secret message transmission. On top of this backbone, several networking layers are implemented, in order to transform the grid of point-to-point connections into a network that allows a user to transmit a secret message between two distant points, not necessarily directly connected by a QKD link, with unconditional security. On the uppermost layer of the network, practical applications such as secure phone communications or private bank transactions, can be implemented.

In order to meet the requirements imposed by such an architecture, and to ensure long-term system operation without human intervention, several features and verifications were implemented in the prototype software. In particular, as explained in detail in~\cite{lodewyck:pra2007}, there is a non-negligible probability that the LDPC codes leave a certain number of errors in the final sequence. Most of the time, these can be deterministically corrected using a BCH (Bose -- Ray-Chaudhuri and Hocquenghem) code~\cite{bose,hocquenghem}. However, if too many errors remain, the BCH code is unable to correct them, and thus Alice and Bob's data are different. Now, in the case of a network such as the SECOQC one, providing the upper network layers with slightly different keys results in loss of synchronization and ultimately in a transmission halt. It is therefore necessary to perform key verification, after the privacy amplification process. The most important property of the hash functions used in privacy amplification is that a single difference in two input bit sequences results in completely different output keys, but two identical input sequences always yield the same output. Hence, to verify the sequences, we compare a random sample of length $n$ of the final keys of Alice and Bob. The probability of failure of this verification process, that is the probability of using an identical sample while the keys are different, is of the order of $2^{-n}$. In our case, $n = 200$ is chosen, which yields a probability of error of $10^{-60}$.

\section{Results and discussion}
\label{sec:results}

The experiments that we present in this section took place during the field implementation of the SECOQC quantum cryptography network in Vienna, in October 2008. A first prototype based on the setup presented in~\cite{lodewyck:pra2007} was installed in Vienna in April 2008, and was then replaced by the advanced prototype presented in this work.

\subsection{System calibration}
\label{sec:labo}

Before performing the QKD experiments, it is essential to perform system calibration. More specifically, there are three basic parameters, expressed in shot-noise units, that determine the secret information rate: the modulation variance $V_\subs{A}$ (at the input of the channel), the variance $V_\subs{B}$ of the data measured by Bob, and the correlation $\rho^2 = \frac{\langle X_\subs{A} X_\subs{B}\rangle^2}{V_\subs{A} V_\subs{B}}$ between the data of Alice and Bob. The value of the correlation $\rho^2$ is independent of any calibration. In order to determine $V_\subs{A}$ and $V_\subs{B}$ from the accessible experimental variances $V_\subs{A}N_0$ and $V_\subs{B}N_0$, it is necessary to calibrate the value of the shot noise, $N_0$. This is achieved by measuring the variance of Bob's data when only the local oscillator is present, i.e. $N = N_0 + \Velec$, and by subtracting the electronic noise variance $\Velec$ measured when neither the signal nor the LO are present. In addition to $N_0$ and $\Velec$, it is also necessary to determine the transmission efficiency of Bob's apparatus, $\eta$. Following the described process, the calibration of Bob's apparatus yields a transmission efficiency $\eta = 0.6$ and an electronic noise $\Velec N_0 = 0.01 N_0$. Furthermore, when imposing a modulation variance $V_\subs{A}N_0 = 10 N_0$, the measured excess noise is typically $0 \leq \varepsilon N_0 < 0.01 N_0$, depending on the surrounding noise and vibration conditions.

After the calibration procedure, the setups of Alice and Bob are placed at each side of a preinstalled optical fibre featuring a transmission efficiency $T = 0.51$. This efficiency corresponds to a standard 15~km optical fibre. In practice, the optical fibre used by the continuous-variable QKD prototype in the SECOQC quantum network was particularly lossy and its length was 9~km.

Based on the above measured parameters, the expected secret key generation rate with a reconciliation efficiency $\beta = 1$, calculated using equation~\ref{eq:rate} and with an optical emission rate of 500~kHz, is $\Delta I \approx$~100 kbit/s.

\subsection{Prototype stability}
\label{sec:stability}

During the six months that the prototypes were in operation several stability tests were made possible. The optoelectronic part of the prototypes did not require any human intervention; in particular, the optical pulse generation system and the detection system were in continuous operation during the entire period at a rate of 500 kHz without downtime. Furthermore, the birefringence, and consequently the polarization, in the installed fibre typically varied ten times slower than a laboratory fibre spool of equivalent length, facilitating the compensation of drifts of this nature.

On the contrary, phase drifts that occur because of temperature changes or vibrations in the environment can lead to important instability of the system. Indeed, as shown in figure~\ref{fig:setup}, the entire device is a Mach-Zehnder interferometre, in which the two paths are separated just after the laser diode and interfere in the homodyne detector. In the setups of both Alice and Bob, the LO and signal paths are separated by 80~m, which leads to relative phase drifts over time.

These phase drifts do not have a noticeable effect on the excess noise as long as their typical variation time is long relatively to the length of the data block used to evaluate the phase. This block is composed of 50000 data points, emitted in 100 ms, while the phase drift linked to temperature changes in the devices is typically of 2$\pi$ every 30 seconds. The effect of temperature on the excess noise is therefore small. Moreover, this drift is almost linear over 1 second, which makes possible an almost perfect control of the phase per pulse, by imposing an opposite linear phase ramp on Bob's phase modulator. The effect of vibrations of the racks containing the devices due to the surrounding environment, however, is more difficult to control. These vibrations have a typical frequency of 50 to 1000 Hz, which is fast compared to the phase evaluation frequency. Since these rapid phase variations cannot be modeled or controlled, it is necessary to isolate the optical devices mechanically. Anti-vibration mountings were placed between the racks and the metal plate supporting the optics to improve the mechanical isolation; nevertheless, the measured excess noise can reach 0.1 $N_0$ in the worst conditions.

Finally, concerning the software part of the prototype, and in particular the classical algorithms and network interface, significant combined efforts of the network developing team of the SECOQC project and the QKD teams participating in the network, led to good stability as well as compatibility of all the systems with the central network managing program.

\subsection{Prototype performance during the SECOQC QKD network implementation}
\label{sec:time}

\begin{figure}[t]
  \centering
  \includegraphics[width=.8\columnwidth]{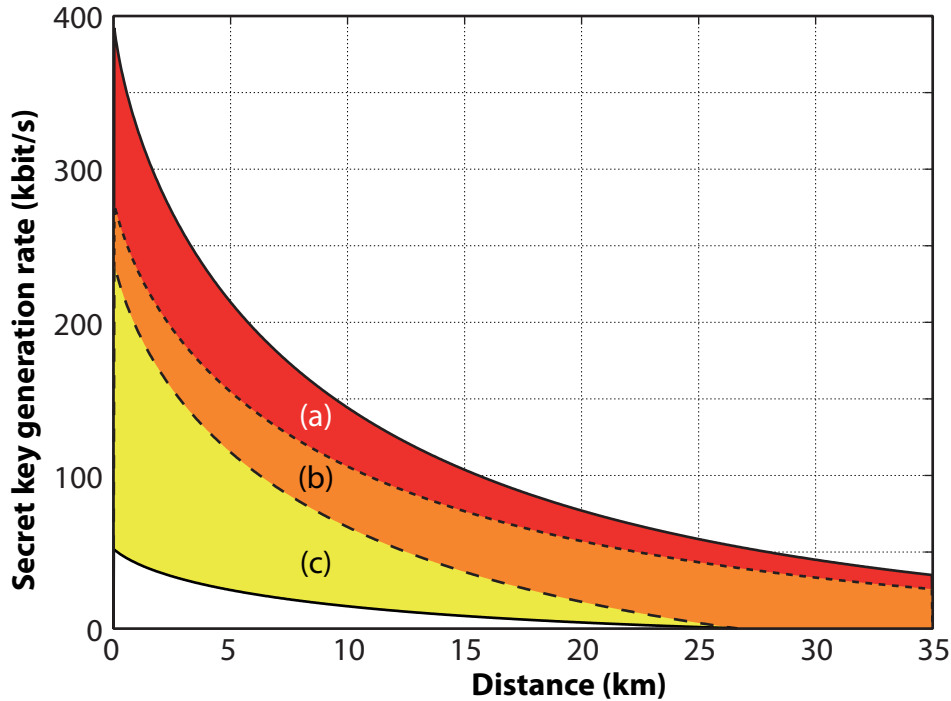}
\caption{Summary of the practical reasons for the decrease of the secret key generation rate. The upper solid curve represents the maximum theoretical rate attainable by the system. (a) Drop in rate due to a realistic excess noise of 4~\%. (b) Drop due to a limited reconciliation efficiency of 90~\%. (c) Drop due to the current impossibility to postprocess all the data at the optical emission rate. The lower solid curve is the practical achievable secret key distribution rate. The curves are drawn using the standard optical fibre loss coefficient of 0.2~dB/km.}
\label{fig:rateloss}
\end{figure}

As we mentioned in section~\ref{sec:labo}, the theoretically expected secret key generation rate of the CVQKD prototype, given the measured system parameters, is 100~kbit/s. However, the actual rate produced by the system during the QKD experiments that we performed was one order of magnitude lower than this value. There are several reasons why this decrease occurs, and it is actually very important to take into account such considerations when designing a practical QKD system for use in real networks. Figure~\ref{fig:rateloss} summarizes the various key drops that occur in the system, which are detailed below.

In most discrete-variable QKD systems, the secret key generation rate is limited by the detector technology, especially regarding dark noise and maximum detection rate. In CVQKD, the optical and optoelectronic components are not the limiting factor; in fact, all the optical components (including the photodiodes) would be able to operate at an optical rate increased by a factor of 100, which would be the case for example if 5 ns pulses emitted at a rate of 50 MHz were used. The limitations of the CVQKD system in terms of secret key generation rate are mainly due to insufficient computational speed and to the effects of environmental perturbations.

\begin{figure}[t]
\centering
  \includegraphics[width=.6\columnwidth]{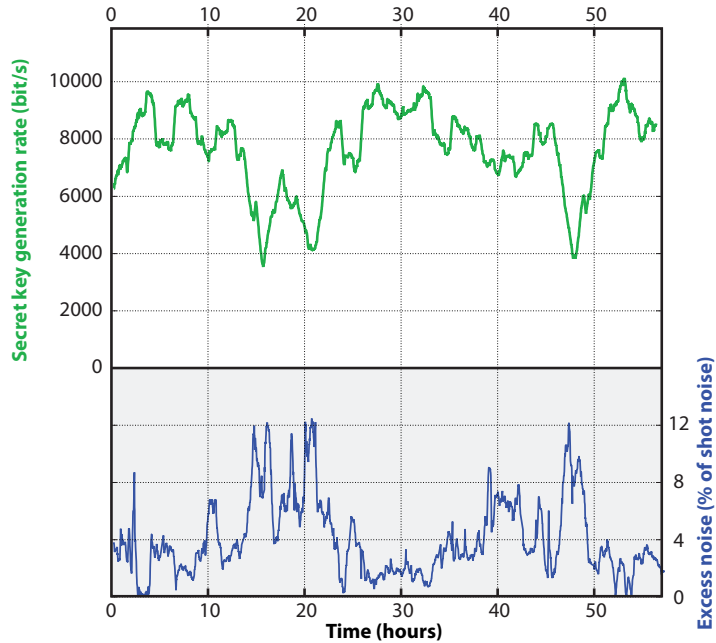}
\caption{Secret key generation rate and excess noise of the system as a function of time. Both curves represent floating averages: over 100 successive keys for the excess noise, and over 1 hour for the secret key rates. This averaging explains the differences between the theoretical key rate shown in figure~\ref{fig:rateexcess} and the actual key rate shown here.}
\label{fig:excess}
\end{figure}

\begin{figure}[t]
  \centering
  \includegraphics[width=.6\columnwidth]{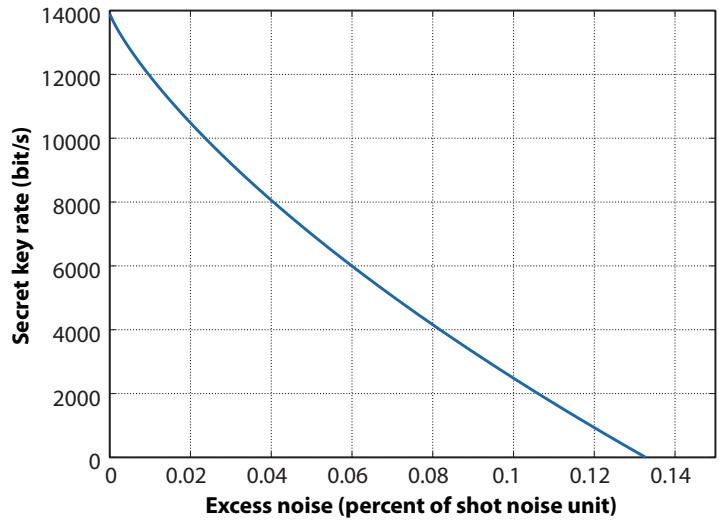}
\caption{Secret key generation rate as a function of excess noise, with all the other parameters fixed as in section~\ref{sec:labo}.}
\label{fig:rateexcess}
\end{figure}

More specifically, the first reason for the key rate drop (part (a) of figure~\ref{fig:rateloss}) is the excess noise induced by the vibrational environment of the prototype. As we observe in figure~\ref{fig:excess}, which shows the results for the secret key generation rate and the excess noise as a function of time during the CVQKD prototype field test, the excess noise varies considerably with time, typically between 0 and 10 \% of the shot noise. This variation has a direct effect on the secret information rate, as illustrated in figure~\ref{fig:rateexcess}.

The second reason for the key rate decrease (part (b) of figure~\ref{fig:rateloss}) is the finite efficiency of the reconciliation phase of the protocol. As we mentioned in section~\ref{sec:scheme}, the information $I_\subs{AB}$ available in the data cannot in practice be entirely extracted, and this effect is taken into account by introducing the parameter $\beta$ that appears in equation~\ref{eq:rate}. In the prototype that we have implemented, $\beta = 0.9$. It is important to note that the finite reconciliation efficiency is also directly linked to the limited transmission distance of the prototype: finding ways of increasing the efficiency of the reconciliation process can result in a significant increase in its range~\cite{leverrier:pra2008}.

Another major decrease in the secret key generation rate (part (c) of figure~\ref{fig:rateloss}) stems from the limitations in computational speed of the processor used for data post-processing. Indeed, the reconciliation and privacy amplification algorithms are complex and very demanding in terms of computational power. The speed optimization of these algorithms depends on many parameters; in particular, there is a trade-off concerning the length of the set of data used to evaluate the parameters: a long set of data decreases the statistical uncertainty on the parameter evaluation, but increases the time required to process the data. For the field test, we chose a length of 2 million pulses per key.  The optical emission of this set of pulses takes 5 seconds with the current optical pulse rate, while the reconciliation of the data (4 million bits) typically lasts 24 seconds and the privacy amplification algorithm another 5 seconds. As a direct consequence, when the system works continuously, one processor core manages to perform the classical post-processing on less than 20\% of the available pulses, which directly results in a loss of a factor of 6 between the optically available secret key rate and the final extracted rate. By using a quadruple-core processor, we have been able to implement three post-processing operations in parallel, which resulted in an increase of the secret key rate by a 2.1 factor, however even with this improvement the processing rate is still 3 times slower than the optical rate.

Finally, the classical communication that is necessary to transmit the various samples needed for channel evaluation and the side information required for reconciliation has a significant effect, which is dependent on the protocol and the implementation of these procedures. As shown in figure~\ref{fig:Rec}, the size of the transmitted messages is quite important in our case, and adds 5 seconds to the key extraction time. On the other hand, most two-way reconciliation algorithms (such as CASCADE) require numerous transmissions of small messages; the message transmission in itself is therefore negligible but the accumulated latency of all the transmissions can lead to important delays. In addition to these delays, a certain amount of key material is used for authentication purposes. In our case, the authentication requires typically 128 bits for every key generated ($\approx$~150\,000 bits), which is negligible.

In the field test, given all the factors described above, the CVQKD prototype generated secret keys during 57 hours with an average rate of 8 kbit/s, as shown in figure~\ref{fig:excess}. From figure~\ref{fig:rateloss} we can see that the maximal attainable communication distance with the current parameters is 27 km, and that the rate increases significantly when the transmission loss decreases. The prototype is therefore particularly adapted to metropolitan communications (up to 20 km) with high speed requirements.

\subsection{Discussion and future developments}
\label{sec:discussion}

The presented results clearly show that developing a table-top laboratory QKD system is a very different procedure than developing a prototype for integration into a realistic environment. Since one of the basic rules of QKD is to systematically attribute the practical imperfections to an eavesdropper's action, stability is of utmost importance, and the operational electromagnetic, vibrational or luminous environment can have a dramatic effect on system performance. Furthermore, the effects related to software cannot be neglected, since QKD systems are designed to work in combination with encrypters and network components, and interactions with such systems often lead to latencies or instabilities.

In the case of the CVQKD prototype, the two main factors that lead to a suboptimal system performance, namely vibrations and computer processor limitations, can certainly be eliminated. In order to avoid all mechanical-induced effects on the phase of the interferometre, the apparatuses should be specifically designed to prevent fibres from vibrating. Concerning the post-processing speed, the system performance is clearly directly affected by the steady increase in computer performance, but optimization work is still possible to achieve faster and more efficient reconciliation. In particular, the massive computing parallelization made possible by new-generation graphical processors is particularly adapted to LDPC-based algorithms. The implementation of such algorithms on graphical processors therefore offers a promising research direction. In parallel with these developments, new algorithms, which aim at reaching longer communication distances and higher secret key generation rates based on more efficient reconciliation, are currently being designed~\cite{leverrier:pra2008}.

\section{Conclusion}
\label{sec:conclusion}

The continuous-variable QKD prototype that we have realized has been integrated into a quantum cryptography telecommunication network, and yielded an average secret key generation rate of 8 kbit/s over a 3 dB loss fibre, during 57 hours. Time and polarization multiplexing have been used to transmit the signal and local oscillator in the same quantum channel, and complex feedback control procedures have been implemented to ensure a stable and automatic operation without human intervention. The system has been proven secure against general coherent eavesdropping attacks. The secret key generation rate is currently mainly limited by the computing possibilities of the system's processor. In parallel with further stabilization of the interferometre's phase, work is currently in progress to increase the speed of classical post-processing. This will allow an increase in the pulse generation rate, and therefore a significant improvement in the final secret key generation rate. With these improvements, the CVQKD prototype can ultimately provide metropolitan-size networks with secret keys generated at rates greater than 100 kbit/s, an objective that appears to be within reach in the near future.

\ack
Financial support for this work was provided by the Integrated European Project SECOQC (Grant No. IST-2002-506813) and the French National Research Agency Project SEQURE. E.D. acknowledges support from the European Union through a Marie-Curie fellowship and a Marie-Curie reintegration grant.

\end{document}